\documentclass[a4,11pt,reqno]{amsart}
\usepackage[english]{babel}    \usepackage[latin1]{inputenc}   \usepackage[T1]{fontenc}     \usepackage[french]{minitoc}
\usepackage[nice]{nicefrac}    \usepackage{latexsym,amsfonts}  
\usepackage{graphics}    \usepackage{ulem}       \usepackage{hhline}    \usepackage{dsfont}    \usepackage{mathrsfs}
\usepackage{fancyhdr}    \usepackage{amsmath}    \usepackage{amssymb}   \usepackage{rotating}  \usepackage{fancybox}
\usepackage{color}       \usepackage{colortbl}   \usepackage{setspace}  \usepackage{enumerate} \usepackage{amsthm}
\usepackage{multicol}    
\usepackage{amsthm}    \usepackage{varioref}  \usepackage{textcomp}
\usepackage{lmodern}     \usepackage{mathpazo}   \usepackage{euscript}  \usepackage[pdftex]{hyperref}
\numberwithin{equation}{section}  \makeatletter\@addtoreset{equation}{section}
   \DeclareMathSymbol{\subsetneqq}{\mathbin}{AMSb}{36}
\newtheorem {theorem}{Theorem}[section]            
\newtheorem {definition}[theorem]{Definition}   \newtheorem {corollary}[theorem]{Corollary}     \newtheorem {remark}[theorem]{Remark}
\newtheorem {proposition}[theorem]{Proposition}       
         
     \newcommand{\fin}{\hfill $\blacksquare$}

\begin{document}
\title[] {\textsf{Husimi's Q-function of the isotonic oscillator in a generalized
negative binomial states representation}}
\author[Z. Mouayn]{{\bf  Zouha\"{i}r MOUAYN }}
\address{ Faculty of Sciences \& Technics (M'Ghila)
          \newline
         BP.523, B\'{e}ni Mellal, Morocco}
\date{\today}
\maketitle

\begin{abstract}
While considering a class of generalized negative binomial states, we verify
that the basic minimum properties for these states to be considered as
coherent states are satisfied. We particularize them for the case of the
Hamiltonian of the isotonic oscillator and we determine the corresponding
Husimi's Q-function. This function may be used to determine a lower bound
for the thermodynamical potential of the Hamiltonian by applying a
Berezin-Lieb inequality.
\end{abstract}

\section{Introduction}

The negative binomial states (NBS) are the field states that are
superposition of the number states with appropriately chosen coefficients \cite{1}. 
Precisely, these labeling coefficients are such that the
associated photon-counting distribution is a negative binomial probability
distribution \cite{2}. As matter of fact, these coefficients
turn out to be an orthonormal basis of a weighted Bergman space of analytic
functions on the complex unit disk satisfying a certain growth condition.
Furthermore, the NBS are considered as intermediate states between pure
coherent states and pure thermal states \cite{3} and reduce to
Susskind-Glogower phases states for a particular limit\thinspace of the
parameter \cite{4}.

Now, as in \cite{5,6}, we replace the labeling coefficients by an
orthonormal basis of a Hilbert space that generalize the weighted Bergman
space of analytic functions on the unit disk we have mentioned to consider a
class of generalized negative binomial states (GNBSs) in this sense. Here,
we precisely verify that the basic minimum properties for the constructed
states to be considered as coherent states are satisfied. Namely, the
conditions which have been formulated by Klauder \cite{7}: $\left(
a\right) $ the continuity of labeling, $\left( b\right) $ the fact that
these states are normalizable but not orthogonal and $\left( c\right) $
these states fulfilled the resolution of the identity with a positive weight
function.

Next, we particularize the GNBSs formalism for the case of the isotonic
oscillator (IO) \cite{8} whose importance consists in the fact
that it admits exact analytic solutions an being in a certain sense an
intermediate potential between the three dimensional harmonic oscillator
potential and other anharmonic potentials such as P\"{o}schl-Teller or Morse
potentials \cite{9}. Next, we determine the Husimi's $Q$-function \cite{10} 
which turns out to be the expectation value in the GNBSs
representation of the heat semigroup operator associated with the IO. The
obtained $Q$-function can be considered as a lower symbol for this heat
semigroup operator. Finally, a lower bound for the thermodynamical potential
of the IO may be obtained by applying a Berezin-Lieb inequality \cite{11,12}.

The paper is organized as follows. In Section 2, we review briefly the
coherent states formalism we will be using. Section 3 deals with some needed
facts on the generalized weighted Bergman spaces on the disk. In Section 4,
we attach to each of these spaces a set of coherent states generalizing the
negative binomial states and verify that they satisfy the basic minimum
properties of coherent states. In Section 5, we recall briefly some needed
spectral properties of the isotonic oscillator Hamiltonian . In Section 6,
we obtain the Husimi's Q-function associated with the IO in the coherent
state representation and we deduce a lower bound for the thermodynamical
potential of the IO.

\section{Coherent states and Berezin-Lieb inequalities}

\noindent Here, we review a coherent states formalism starting from a
measure space ''{it as a set of data}'' as presented in \cite{13}. 
Let $X=\left\{ x\mid x\in X\right\} $ be a set equipped with a measure $%
d\mu $ and $L^{2}(X,d\mu )$\ the space of $d\mu$-square integrable
functions on $X$. Let $\mathcal{A}^{2}\subset L^{2}(X,d\mu )$ be a subspace
of infinite dimension with an orthonormal basis $\left\{ \Phi _{j}\right\}
_{j=0}^{\infty }$. Let $\mathcal{H}$ be another (functional) space with $%
\dim \mathcal{H}=\infty $ and $\left\{ \phi _{j}\right\} _{j=1}^{\infty }$
is a given orthonormal basis of $\mathcal{H}$. Then consider the family of
states $\left\{ \mid x>\right\} _{x\in X}$ in $\mathcal{H}$, through the
following linear superpositions:
\begin{equation}
\mid x>:=\left( \mathcal{N}\left( x\right) \right) ^{-\frac{1}{2}%
}\sum_{j=0}^{+\infty }\Phi _{j}\left( x\right) \mid \phi _{j}>\quad
\label{2.1}
\end{equation}
where
\begin{equation}
\mathcal{N}\left( x\right) =\sum_{j=0}^{+\infty }\Phi _{j}\left( x\right)
\overline{\Phi _{j}\left( x\right) }.  \label{2.2}
\end{equation}
These coherent states obey the normalization condition
\begin{equation}
\left\langle x\mid x\right\rangle _{\mathcal{H}}=1  \label{2.3}
\end{equation}
and the following resolution of the identity of $\mathcal{H}$%
\begin{equation}
\mathbf{1}_{\mathcal{H}}=\int _{X}\mid x><x\mid \mathcal{N}\left(
x\right) d\mu \left( x\right)   \label{2.4}
\end{equation}
which is expressed in terms of Dirac's bra-ket notation $\mid x><x\mid $
meaning the rank-one -operator $\varphi \mapsto \left\langle \varphi \mid
x\right\rangle _{\mathcal{H}}.\mid x>$. \ The choice of the Hilbert space $%
\mathcal{H}$ define in fact a quantization of the space $X$ \ by the
coherent states in \eqref{2.1}, via the inclusion map $X\ni
x\mapsto \mid x>\in \mathcal{H}$ and the property \eqref{2.4} is
crucial in setting the bridge between the classical and the quantum mechanics.

Now, given a set of coherent states $\left\{ \mid x>\right\} $, the concept
of upper and lower symbols of an operator $A$  was separately introduced by
Berezin \cite{11} and Lieb \cite{12} by
\begin{equation}
A=\int _{X}d\mu \left( x\right) \widehat{A}\mid x><x\mid   \label{2.5}
\end{equation}
to define the upper symbol $\widehat{A}$ of $A,$ and the expectation value
\begin{equation}
\widetilde{A}\left( x\right) :=\left\langle x\mid A\mid x\right\rangle
\label{2.6}
\end{equation}
for the definition of the lower symbol $\widetilde{A}$ of $A.$ Note that
given an operator $A$ its upper symbol is not unique in general. It can be
proved \cite{12} that given any convex function $\phi ,$ the
following inequalities
\begin{equation}
\int _{X}\phi \left( \widetilde{A}\right) d\mu \left( x\right) \leq
Tr\left( \phi \left( A\right) \right) \leq \int _{X}\phi \left(
\widehat{A}\right) d\mu \left( x\right)   \label{2.7}
\end{equation}
hold and are called Berezin-Lieb inequalities.

\section{Generalized Bergman spaces on $\mathbb{D}$}

Let $\mathbb{D}=\left\{ z\in \mathbb{C},\left| z\right| <1\right\} $ be unit
disk endowed with its usual Kh\"{a}ler metric $ds^{2}=-\partial \overline{%
\partial }Log\left( 1-z\overline{z}\right) dz\otimes d\overline{z}.$ The
Bergman distance on $\mathbb{D}$ is given by
\begin{equation}
\cosh ^{2}d\left( z,w\right) =\frac{(1-z\overline{w})(1-\overline{z}w)}{%
\left( 1-z\overline{z}\right) \left( 1-w\overline{w}\right) }  \label{3.1}
\end{equation}
and the volume element reads
\begin{equation}
d\mu \left( z\right) =\frac{1}{\left( 1-z\overline{z}\right) ^{2}}d\nu
\left( z\right)   \label{3.2}
\end{equation}
with the Lebesgue measure $d\nu \left( z\right) .$ Let us consider the $1-$%
form on $\mathbb{D}$ defined by $\theta =-i\left( \partial -\overline{%
\partial }\right) Log\left( 1-z\overline{z}\right) $ to which the Schr\"{o}%
dinger operator
\begin{equation}
H_{\sigma }:=\left( d+i\frac{\sigma }{2}ext\left( \theta \right) \right)
^{\ast }\left( d+i\frac{\sigma }{2}ext\left( \theta \right) \right)
\label{3.3}
\end{equation}
can be associated. Here $\sigma \geq 0$ is a fixed number, $d$ denotes the
usual exterior derivative on differential forms on $\mathbb{D}$ and $%
ext\left( \theta \right) $ is the exterior multiplication by $\theta $ while
the symbol $\ast $ stands for the adjoint operator with respect to the
Hermitian scalar product induced by the Bergman metric $ds^{2}$ on
differential forms. Actually, the operator $H_{\sigma }$ is acting on the
Hilbert space $L^{2}\left( \mathbb{D},d\mu \left( z\right) \right) $ and can
be unitarly intertwined as
\begin{equation}
\left( 1-z\overline{z}\right) ^{\frac{1}{2}\sigma }\Delta _{\sigma }\left(
1-z\overline{z}\right) ^{-\frac{1}{2}\sigma }=H_{\sigma }  \label{3.4}
\end{equation}
in terms of the second order differential operator
\begin{equation}
\Delta _{\sigma }:=-4\left( 1-z\overline{z}\right) \left( \left( 1-z%
\overline{z}\right) \frac{\partial ^{2}}{\partial z\partial \overline{z}}%
-\sigma \overline{z}\frac{\partial }{\partial \overline{z}}\right) .
\label{3.5}
\end{equation}
The latter one is acting on the Hilbert space
\begin{equation}
L^{2,\sigma }\left( \mathbb{D}\right) =L^{2}\left( \mathbb{D},\left( 1-z%
\overline{z}\right) ^{\sigma -2}d\nu \left( z\right) \right) .  \label{3.6}
\end{equation}
The spectral analysis of $\Delta _{\sigma }$ have been studied by many
authors, see \cite{5} and references therein. Note that this
operator is an elliptic densely defined operator on $L^{2,\sigma }\left(
\mathbb{D}\right) $ and admits a unique self-adjoint realization that we
denote also by $\Delta _{\sigma }.$ The part of its spectrum is not empty if
and only if $\sigma >1.$ This discrete part consists of eigenvalues
occurring with infinite multiplicities and having the expression
\begin{equation}
\epsilon _{m}^{\sigma }:=4m\left( \sigma -m-1\right)   \label{3.7}
\end{equation}
for varying $m=0,1,...,\left[ (\sigma -1)/2\right] .$\ Here, $[x]$ denotes the greatest integer not exceeding $x.$ Moreover, it is well known
that the functions given in terms of Jacobi polynomials \cite{15,17}
 by
\begin{align}
\Phi _{k}^{\sigma ,m}\left( z\right)  = &\sqrt{\frac{\left( \sigma -2m-1\right)
k!\Gamma \left( \sigma -m\right) }{\pi m!\Gamma \left( \sigma -2m+k\right) }} \label{3.8} \\ 
& \times \left( -1\right) ^{k}\left( 1-z\overline{z}\right) ^{-m}\overline{z}%
^{m-k}P_{k}^{\left( m-k,\sigma -2m-1\right) }\left( 1-2z\overline{z}\right)
\nonumber
\end{align}
constitute an orthonormal basis of the eigenspace
\begin{equation}
\mathcal{A}_{m}^{2,\sigma }\left( \mathbb{D}\right) :=\left\{ \phi \in
L^{2,\sigma }\left( \mathbb{D}\right) ,\Delta _{\sigma }\phi =\epsilon
_{m}^{\sigma }\phi \right\} .  \label{3.9}
\end{equation}
of $\Delta _{\sigma }$ associated with the eigenvalue\ $\epsilon
_{m}^{\sigma }$ in \eqref{3.7}. Finally, the $L^{2}-$eigenspace $\mathcal{A}_{0}^{2,\sigma }\left( \mathbb{D}\right) =\left\{ \phi \in
L^{2,\sigma }\left( \mathbb{D}\right) ,\Delta _{\sigma }\phi =0\right\} $
corresponding to $m=0$ and associated to $\epsilon _{0}^{\sigma }=0$ in \eqref{3.7} reduces further to the weighted Bergman space consisting
of holomorphic functions $\phi $ : $\mathbb{D\rightarrow C}$ with the growth
condition
\begin{equation}
\int _{\mathbb{D}}\left| \phi \left( z\right) \right| ^{2}\left( 1-z\overline{z}\right) ^{\sigma -2}d\nu \left( z\right) <+\infty .  \label{3.10}
\end{equation}
This is why the eigenspaces in \eqref{3.9} have been called
generalized Bergman spaces on the complex unit disk.

\begin{remark}\label{Rem3.1} In \cite{5}, we have used the basis in \eqref{3.8} to perform a class of coherent states belonging to the
Hilbert space $L^{2}\left( \mathbb{R}_{+}^{\ast },x^{-1}dx\right) $. The
associated coherent state transform have been considered as a generalization
of the second Bargmann transform (\cite[p.203]{18}).
\end{remark}

\section{Generalized negative binomial states}

\noindent The negative binomial states are labeled by points $z \in
\mathbb{D}$ and are of the form
\begin{equation}
\mid z\text{ },\sigma ,0>:=\left( 1-z\overline{z}\right) ^{\frac{1}{2}\sigma
}\sum\limits_{k=0}^{+\infty }\sqrt{\frac{\Gamma \left( \sigma +k\right) }{
\Gamma \left( \sigma \right) k!}}z^{k}\mid \psi _{k}>,  \label{4.1}
\end{equation}
where $\sigma >1$ is a fixed parameter and $\mid \psi _{k}>$ are Fock
states. Their photon probability distribution
\begin{equation}
\left| <\psi _{k}\mid z\text{ },\sigma >\right| ^{2}=\left( 1-z\overline{z}%
\right) ^{\sigma }\left( z\overline{z}\right) ^{k}\frac{\Gamma \left( \sigma
+k\right) }{\Gamma \left( \sigma \right) k!}  \label{4.2}
\end{equation}
obeys the negative binomial probability distribution with parameters $%
\lambda =z\overline{z}$ and $\sigma .$ Also, observe that the coefficients
in the superposition \eqref{4.1}:
\begin{equation}
\Phi _{k}^{\sigma ,0}\left( z\right) :=\sqrt{\frac{\Gamma \left( \sigma
+k\right) }{\pi \Gamma \left( \sigma \right) k!}}z^{k},k=0,1,2,...  \label{4.3}
\end{equation}
constitute an orthonormal basis of the eigenspace $\mathcal{A}_{0}^{2,\sigma
}\left( \mathbb{D}\right) $ associated with the first eigenvalue $\epsilon
_{0}^{\sigma }=0$ and consisting of analytic functions on $\mathbb{D}$ with
the growth condition \eqref{3.10}.

For instance, let $\sigma >1$ and $m=0,1,...,\left[ \left( \sigma -1\right)
/2\right]$ be fixed parameters and let $\left\{ \mid \psi
_{k}>\right\} _{k=0}^{\infty }$\ be a set of Fock states in a Hilbert space $%
\mathcal{H}$.  Then a class of generalized negative binomial states (GNBS)
can be defined as in \cite{5,6} by
\begin{equation}
\mid z\text{ },\sigma ,m>=\left( \mathcal{N}_{\sigma ,m}\left( z\right)
\right) ^{-\frac{1}{2}}\sum\limits_{k=0}^{+\infty }\Phi _{k}^{\sigma
,m}\left( z\right) \mid \psi _{k}> , \label{4.4}
\end{equation}
where $\mathcal{N}_{\sigma ,m}\left( z\right) $\ is a normalization factor
and $\left\{ \Phi _{k}^{\sigma ,m}\left( z\right) \right\} _{k=0}^{\infty }$%
\ \ is the orthonormal basis of the generalized Bergman space in \eqref{3.9}. Now, one of the important task to do is to determine is the
overlap relation between two GNBSs.

\begin{proposition}\label{Prop4.1} Let $\sigma >1$ and $m=0,1,...,
\left[ \left( \sigma -1\right) /2\right] .$ Then, for every $z,w\in
\mathbb{D}$, the overlap relation between two GNBSs is given through
the scalar product
\begin{align}
&<w,\sigma ,m\mid z\text{ },\sigma ,m>_{\mathcal{H}} 
=\frac{(\sigma
-2m-1)\Gamma \left( \sigma -m\right) \left( \mathcal{N}\left( z\right)
\mathcal{N}\left( w\right) \right) ^{-\frac{1}{2}}}{\pi m!\left( -1\right)
^{m}\Gamma \left( \sigma -2m\right) \left( 1-z\overline{w}\right) ^{\sigma }}
\label{4.5}
\\
&\times \left( \frac{(1-z\overline{w})\left( 1-\overline{w}z\right) }{\left(
1-z\overline{z}\right) \left( 1-w\overline{w}\right) }\right)
^{m}._{2}\digamma _{1}\left( -m,\sigma -m,\sigma -2m;\frac{\left( 1-z%
\overline{z}\right) \left( 1-w\overline{w}\right) }{(1-z\overline{w})\left(
1-\overline{w}z\right) }\right), \nonumber
\end{align}
where $_{2}\digamma _{1}$  is a terminating Gauss
hypergeometric sum. 
\end{proposition}

\noindent\textbf{Proof. }In view of Eq. \eqref{4.4}, the scalar product of
two GNBS $\mid z$ $,\sigma ,m>$ and $\mid w$ $,\sigma ,m>$ in $\mathcal{H}$
reads
\begin{align}
<w,\sigma ,m\mid z\text{ },\sigma ,m>_{\mathcal{H}}&=\left( \mathcal{N}%
_{\sigma ,m}\left( z\right) \mathcal{N}_{\sigma ,m}\left( w\right) \right)
^{-\frac{1}{2}}\sum\limits_{k=0}^{+\infty }\Phi _{k}^{\sigma ,m}\left(
z\right) \overline{\Phi _{k}^{\sigma ,m}\left( w\right) }
\nonumber \\
&=\frac{(\sigma -2m-1)\left( \mathcal{N}\left( z\right) \mathcal{N}\left(
w\right) \right) ^{-\frac{1}{2}}}{\pi \left( \left( 1-z\overline{z}\right)
\left( 1-w\overline{w}\right) \right) ^{m}}\mathcal{S}_{z,w}^{\sigma ,m},
\label{4.6}
\end{align}
where
\begin{align}
\mathcal{S}_{z,w}^{\sigma ,m}&=\frac{\Gamma \left( \sigma -m\right) \left(
\overline{z}w\right) ^{m}}{m!\Gamma \left( \sigma -2m\right) }%
\sum_{k=0}^{+\infty }\frac{k!}{(\sigma -2m)_{k}}\left( \frac{1}{\overline{z}w%
}\right) ^{k}  \label{4.7}
\\
&\times P_{k}^{\left( m-k,\sigma -2m-1\right) }\left( 1-2z\overline{z}\right)
P_{k}^{\left( m-k,\sigma -2m-1\right) }\left( 1-2w\overline{w}\right) .
\nonumber
\end{align}
Making use of the following identity (\cite[p.1329]{19}):
\begin{align}
&\sum_{n=0}^{+\infty }\frac{n!t^{n}}{\left( 1+\alpha \right) _{n}}
 P_{n}^{(\gamma -n,\alpha )}\left( x\right) P_{n}^{(\gamma -n,\alpha )}\left(
y\right) =\left( 1-\frac{1}{4}\left( x-1\right) \left( y-1\right) t\right)
^{1+\gamma +\alpha }
 \label{4.8}  \\
& \times \left( 1-t\right) ^{\gamma }._{2}\digamma _{1}\left( 1+\gamma +\alpha
,-\gamma ,1+\alpha ;\frac{-\left( x+1\right) \left( y+1\right) t}{\left(
1-t\right) \left( 4-\left( x-1\right) \left( y-1\right) t\right) }\right)
\nonumber
\end{align}
for $n=k,t=1/\overline{z}w,\gamma =m,\alpha =\sigma -2m-1,x=1-2z\overline{z}$
and $y=1-2w\overline{w},$ we obtain, after calculations, the expression
\begin{align}
\mathcal{S}_{z,w}^{\sigma ,m}&=\frac{\Gamma \left( \sigma -m\right) \left(
-1\right) ^{m}}{m!\Gamma \left( \sigma -2m\right) }\frac{\left( (1-z%
\overline{w})\left( 1-\overline{w}z\right) \right) ^{m}}{\left( 1-z\overline{%
w}\right) ^{\sigma }}  \label{4.9} \\
&\times _{2}\digamma _{1}\left( -m,\sigma -m,\sigma -2m;\frac{\left( 1-z%
\overline{z}\right) \left( 1-w\overline{w}\right) }{(1-z\overline{w})\left(
1-\overline{w}z\right) }\right) .\nonumber
\end{align}
Returning back to Eq. \eqref{4.6} and inserting the expression \eqref{4.9} we arrive at the announced formula.   \fin

\begin{corollary}\label{Cor4.1} The normalization factor in \eqref{4.4} is given by
\begin{equation}
\mathcal{N}_{\sigma ,m}\left( z\right) =\frac{(\sigma -2m-1)}{\pi \left( 1-z%
\overline{z}\right) ^{\sigma }},  \label{4.10}
\end{equation}
for every $z\in \mathbb{D}.$
\end{corollary}

\noindent \textbf{Proof.} We first make appeal to the relation (\cite[p.212]{17}):
\begin{equation}
_{2}\digamma _{1}\left( -n,n+\kappa +\varrho +1,1+\kappa ;\frac{1-\tau }{2}%
\right) =\frac{n!\Gamma \left( 1+\kappa \right) }{\Gamma \left( 1+\kappa
+n\right) }P_{n}^{\left( \kappa ,\varrho \right) }\left( \tau \right)
\label{4.11}
\end{equation}
connecting the $_{2}\digamma _{1}-$sum with the Jacobi polynomial for the
parameters $n=m,\kappa =\sigma -2m-1,\varrho =0$ and the variable
\begin{equation}
\tau =1-2\frac{\left( 1-z\overline{z}\right) \left( 1-w\overline{w}\right) }{%
(1-z\overline{w})\left( 1-\overline{w}z\right) }  \label{4.12}
\end{equation}
to rewrite Eq. \eqref{4.5} as
\begin{align}
\sqrt{\mathcal{N}_{\sigma ,m}\left( z\right) \mathcal{N}_{\sigma ,m}\left(
w\right) } & =\frac{\left( -1\right) ^{m}(\sigma -2m-1)\left( 1-z\overline{w}%
\right) ^{-\sigma }}{\pi <w;\sigma ,m\mid z\text{ },\sigma ,m>_{\mathcal{H}}} \label{4.13} \\
& \times
P_{m}^{\left( \sigma -2m-1,0\right) }\left( 1-2\frac{\left( 1-z\overline{z}%
\right) \left( 1-w\overline{w}\right) }{(1-z\overline{w})\left( 1-\overline{w%
}z\right) }\right) .  \nonumber
\end{align}
The factor $\mathcal{N}_{\sigma ,m}\left( z\right) $ should be such that
\begin{equation}
<z,\sigma ,m\mid z,\sigma ,m>_{\mathcal{H}}=1.  \label{4.14}
\end{equation}
So that we put $z=w$ in \eqref{4.13} and we use the symmetry
identity (\cite[p.210]{17}):
\begin{equation}
P_{m}^{\left( \gamma ,\varrho \right) }\left( \xi \right) =\left( -1\right)
^{m}P_{m}^{\left( \varrho ,\gamma \right) }\left( -\xi \right)   \label{4.15}
\end{equation}
to obtain the expression
\begin{equation}
\mathcal{N}_{\sigma ,m}\left( z\right) =\frac{(\sigma -2m-1)}{\pi \left( 1-z%
\overline{z}\right) ^{\sigma }}P_{m}^{\left( 0,\sigma -2m-1\right) }\left(
1\right)   \label{4.16}
\end{equation}
Finally, we apply the fact that (\cite[p.1329]{17}):
\begin{equation}
P_{n}^{\left( \alpha ,\varrho \right) }\left( 1\right) =\frac{\Gamma \left(
n+\alpha +1\right) }{n!\Gamma \left( \alpha +1\right) }  \label{4.17}
\end{equation}
in the case of $\alpha =0,n=m$ and $\varrho =\sigma -2m-1$. This ends the
proof.\fin

 \begin{proposition}\label{Prop4.2} Let $\sigma >1$  and $m=0,1,...,%
\left[ \left( \sigma -1\right) /2\right] .$  Then, the GNBS in  
\eqref{4.4} satisfy the following resolution of the identity
\begin{equation}
\mathbf{1}_{\mathcal{H}}=\int _{\mathbb{D}}\mid z,\sigma ,m><z,\sigma
,m\mid d\mu _{\sigma ,m}\left( z\right),  \label{4.18}
\end{equation}
where $1_{\mathcal{H}}$  is the identity operator and $
d\mu _{\sigma ,m}\left( z\right) $  is a measure which can be
expressed through a Meijer's $G$-function as
\begin{equation}
d\mu _{\sigma ,m}\left( z\right) :=\pi ^{-1}(\sigma -2m-1)G_{11}^{11}\left(
-z\overline{z}\mid
\begin{array}{c}
-1 \\
0
\end{array}
\right) d\nu \left( z\right) ,  \label{4.18}
\end{equation}
and $d\nu \left( z\right) $  being the Lebesgue measure on $%
\mathbb{D}$.
 \end{proposition}

\noindent \textbf{Proof}. Let us assume that the measure takes the form
\begin{equation}
d\mu _{\sigma ,m}\left( z\right) =\mathcal{N}_{\sigma ,m}\left( z\right)
\Omega \left( z\right) d\nu \left( z\right)  , \label{4.19}
\end{equation}
where $\Omega \left( z\right) $ is an auxiliary density to be determined.
Let $\varphi \in \mathcal{H}$ and let us start by writing the following
action
\begin{align}
\mathcal{O}\left[ \varphi \right] &:=\left( \int _{\mathbb{D}}\mid
z,\sigma ,m><z,\sigma ,m\mid d\mu _{\sigma ,m}\left( z\right) \right) \left[
\varphi \right]   \label{4.20}
\\
&=\int _{\mathbb{D}}<\varphi \mid z,\sigma ,m><z,\sigma ,m\mid d\mu
_{\sigma ,m}\left( z\right) .  \label{4.21}
\end{align}
Making use Eq. \eqref{4.4}, we obtain successively
\begin{align}
\mathcal{O}\left[ \varphi \right] &=\int _{\mathbb{D}}<\varphi \mid
\left( \mathcal{N}_{\sigma ,m}\left( z\right) \right) ^{-\frac{1}{2}%
}\sum_{k=0}^{+\infty }\Phi _{k}^{\sigma ,m}\left( z\right) \mid \psi
_{k}>><z,\sigma ,m\mid d\mu _{\sigma ,m}\left( z\right)   \label{4.22}
\\
&=\int _{\mathbb{D}}\sum_{k=0}^{+\infty }\Phi _{k}^{\sigma ,m}\left(
z\right) <\varphi \mid \psi _{k}><z,\sigma ,m\mid \left( \mathcal{N}_{\sigma
,m}\left( z\right) \right) ^{-\frac{1}{2}}d\mu _{\sigma ,m}\left( z\right)
\label{4.23}
\\
&=\left( \sum_{j,k=0}^{+\infty }\int _{\mathbb{D}}\overline{\Phi
_{j}^{\sigma ,m}\left( z\right) }\Phi _{k}^{\sigma ,m}\left( z\right) \mid
\psi _{k}><\psi _{j}\mid \left( \mathcal{N}_{\sigma ,m}\left( z\right)
\right) ^{-1}d\mu _{\sigma ,m}\left( z\right) \right) \left[ \varphi \right]
.  \label{4.24}
\end{align}
We replace the measure $d\mu _{\sigma ,m}\left( z\right) $ by the expression
in the right hand side of Eq. \eqref{4.19}, then Eq. \eqref{4.24} can be written without $\varphi $ as follows
\begin{equation}
\mathcal{O}=\sum_{j,k=0}^{+\infty }\left[ \int _{\mathbb{D}}\overline{%
\Phi _{j}^{\sigma ,m}\left( z\right) }\Phi _{k}^{\sigma ,m}\left( z\right)
\Omega \left( z\right) d\nu \left( z\right) \right] \mid \psi _{j}><\psi
_{k}\mid .  \label{4.25}
\end{equation}
Therefore, we need to have
\begin{equation}
\int _{\mathbb{D}}\overline{\Phi _{j}^{\sigma ,m}\left( z\right) }\Phi
_{k}^{\sigma ,m}\left( z\right) \Omega \left( z\right) d\nu \left( z\right)
=\delta _{jk}.  \label{4.26}
\end{equation}
For this we recall the orthogonality relation of the $\Phi _{k}^{\sigma
,m}\left( z\right) $ in the Hilbert space $L^{2,\sigma }\left( \mathbb{D}%
\right) $, which reads
\begin{equation}
\int _{\mathbb{D}}\overline{\Phi _{j}^{\sigma ,m}\left( z\right) }\Phi
_{k}^{\sigma ,m}\left( z\right) \left( 1-z\overline{z}\right) ^{\sigma
-2}d\nu \left( z\right) =\delta _{jk}.  \label{4.27}
\end{equation}
This suggests us to set $\Omega \left( z\right) :=\left( 1-z\overline{z}%
\right) ^{\sigma -2}.$ Therefore, we get that
\begin{equation}
d\mu _{\sigma ,m}\left( z\right) =\frac{(\sigma -2m-1)}{\pi \left( 1-z%
\overline{z}\right) ^{2}}d\nu \left( z\right) .  \label{4.28}
\end{equation}
By making us of the identity \cite{20}:
\begin{equation}
G_{11}^{11}\left( \zeta \mid
\begin{array}{c}
a \\
b
\end{array}
\right) =\Gamma \left( 1-a+b\right) \zeta ^{b}\left( 1+\zeta \right) ^{a-b-1}
\label{4.29}
\end{equation}
for $\zeta =-z\overline{z}$, $a=-1$ and $b=0$, we arrive at the expression
of the measure in \eqref{4.18}. Therefore, Eq. \eqref{4.25} reduces to
\begin{equation}
\mathcal{O}=\sum\limits_{j,k=0}^{+\infty }\delta _{jk}\mid \psi _{j}><\psi
_{k}\mid =\mathbf{1}_{\mathcal{H}}.  \label{4.30}
\end{equation}
The proof is finished. \fin

 \begin{proposition}\label{Prop4.3} Let $\sigma >1$ \textit{and} $m=0,1,...,%
\left[ \left( \sigma -1\right) /2\right] .$ Then, the
states $\mid z,\sigma ,m>$  satisfy the continuity property with
respect to the label $z\in \mathbb{D}.$  That is, the norm of the
difference of two states 
\begin{equation}
d_{\sigma ,m}\left( z,w\right) \mathit{\ :}=\left\| \left( \mid z,\sigma
,m>-\mid w,\sigma ,m>\right) \right\| _{\mathcal{H}}  \label{4.31}
\end{equation}
goes to zero whenever  $z\rightarrow w.$
 \end{proposition}
 
\noindent \textbf{Proof. }By using the fact that any GNBS is normalized by the factor
given in \eqref{4.10}, direct calculations enable us to write the
square of the quantity in \eqref{4.31} as
\begin{equation}
\mathit{\ }d_{\sigma ,m}^{2}\left( z,w\right) \mathit{\ }=2\left( 1-\Re e
<z,\sigma ,m\mid w,\sigma ,m>\right) .  \label{4.32}
\end{equation}
Next, we use of the expression of the scalar product in \eqref{3.9} form which it is clear that the overlap takes the value $1$ as $z\rightarrow
w$ and consequently $d_{\sigma ,m}\left( z,w\right) \rightarrow 0$.  \fin\\

We end this section by the following remarks.

\begin{remark}\label{Rem3.2} By a general fact on reproducing kernels \cite{21},
 the proof of proposition \eqref{2.1} also says that the
knowledge of the explicit orthonormal basis in \eqref{3.8} leads
directly to expression of the reproducing kernel of the generalized Bergman
space in \eqref{3.9} via calculations using the formula \eqref{4.8} due to A. Srivastava and A. B. Rao \cite{19}.
\end{remark}

\begin{remark}\label{Rem3.2} In \cite{6}, we have used the same basis 
\eqref{3.8} under another form as labeling coefficients in order to
consider the photon-counting probability distribution with the mass function
\begin{equation}
p_{k}^{(\sigma ,m,\lambda )}:=\gamma _{\sigma ,m,k}\left( 1-\lambda \right)
^{\sigma -2m}\lambda ^{\left| m-k\right| }\left( P_{\frac{1}{2}\left(
m+k-\left| m-k\right| \right) }^{\left( \left| m-k\right| ,\sigma
-2m-1\right) }\left( 1-2\lambda \right) \right) ^{2} , \label{4.33}
\end{equation}
where $k=0,1,2,...,$ $\gamma _{\sigma ,m,k}>0$ is a constant, $m=0,1,...,%
\left[ \sigma -1)/2\right]$ and $\lambda =z\overline{z}$. We have
calculated the associated Mandel parameter \cite{22} and we have
discussed the classicality/nonclassicality of the GNBS with respect to the
location of their labeling points $z$ inside the hyperbolic disk $\mathbb{D}
$. Similar results, in the Euclidean plane and the Riemann sphere settings
have been obtained respectively in \cite{23} and \cite{24}.
\end{remark}

\begin{remark}\label{Rem3.3} The fact that we have written the measure $d\mu
_{\sigma ,m}\left( z\right) $ in \eqref{4.18} in terms of the
Meijer's G-function could be of help when tackling the ''\textit{%
photon-added coherent states} (\textit{PACS})'' problem for the GNBS under
consideration.
\end{remark}

\section{The isotonic oscillator $\mathbf{L}_{\protect\alpha }$}

Not all quantum Hamiltonians are known to have exact solutions. An important
model of a solvable class is the isotonic oscillator \cite{25}
\begin{equation}
\mathbf{L}_{\alpha }:=\frac{1}{2}\left( -\frac{d^{2}}{dx^{2}}+x^{2}+\frac{%
\left( \alpha ^{2}-1/4\right) }{x^{2}}\right) ,\alpha \geq 1/2  \label{5.1}
\end{equation}
acting in the Hilbert space\ $\mathcal{H}:=L^{2}\left( \mathbb{R}%
_{+},dx\right) $ and the eigenfunctions $\psi \in \mathcal{H}$ satisfy the
Dirichlet boundary condition $\psi \left( 0\right) =0.$ This operator
appears in the literature under many names such as Gol'dman-Krivchenkov
Hamiltonian \cite{26} or pseudoharmonic oscillator \cite{8}
 or Laguerre operator \cite{27}. It is the generalization the
harmonic oscillator in three dimensions where the generalization lies in the
parameter $\nu =\alpha ^{2}-1/4$ ranging over $\left(0,+\infty \right) $
instead of the angular momentum quantum numbers $l=0,1,2,...$ . This
operator may be factorized as follows
\begin{equation}
\mathbf{L}_{\alpha }=D_{\alpha }^{+}D_{\alpha }^{-}+\alpha +1  \label{5.2}
\end{equation}
in terms of the operators, having the form
\begin{equation}
D_{\alpha }^{\pm }=\frac{1}{\sqrt{2}}\left( -\frac{\alpha +1/2}{x}+x\pm
\frac{d}{dx}\right) .  \label{5.3}
\end{equation}
It is well known that the Hamiltonian $\mathbf{L}_{\alpha }$ admits exact
solutions of the form
\begin{equation}
\psi _{k}^{\alpha }\left( x\right) =\left( \frac{2\Gamma \left( k+1\right) }{%
\Gamma \left( k+\alpha +1\right) }\right) ^{\frac{1}{2}}x^{\alpha +\frac{1}{2%
}}e^{-\frac{1}{2}x^{2}}L_{k}^{\left( \alpha \right) }\left( x^{2}\right),
\label{5.4}
\end{equation}
where $L_{k}^{\left( \alpha \right) }\left( .\right) $ denotes the Laguerre
polynomial \cite{16} and the corresponding eigenvalues are given
by (\cite{28}):
\begin{equation}
\lambda _{k}^{\alpha }=2\alpha +k+1,k=0,1,2,...\text{ \ }.  \label{5.5}
\end{equation}
Note also that the functions $\psi _{k}^{\alpha }$ can be obtained by $k-$%
fold application of a creation operator to the ground state wavefunction $%
\psi _{0}^{\alpha }$. The vectors $\left\{ \mid \psi _{k}^{\alpha }>\right\}
_{k=0}^{\infty }$ satisfy the orthogonality relation
\begin{equation}
\left\langle \psi _{k}^{\alpha }\mid \psi _{k}^{\alpha }\right\rangle _{%
\mathcal{H}}=\int _{0}^{+\infty }\psi _{k}^{\alpha }\left( x\right)
\psi _{j}^{\alpha }\left( x\right) dx=\delta _{kj}  \label{5.6}
\end{equation}
and constitute a complete orthonormal basis for the Hilbert space $\mathcal{H%
}$. Furthermore, they can be used together with the eigenvalues in \eqref{5.5} to define the heat semigroup associated with $\mathbf{L}%
_{\alpha }$ as \textit{\ }
\begin{equation}
e^{-t\mathbf{L}_{\alpha }}\left[ f\right] :=\sum\limits_{k=0}^{+\infty
}e^{-t\lambda _{k}^{\alpha }}<f\mid \psi _{k}^{\alpha }>_{\mathcal{H}}.\psi
_{k}^{\alpha }  \label{5.7}
\end{equation}
for any function $f\in L^{2}\left( \mathbb{R}_{+},dx\right) $ .It is also
well known that by using the Hille-Hardy formula (\cite[p.242]{17}), this semigroup has an integral representation, i.e.,
\begin{equation}
e^{-t\mathbf{L}_{\alpha }}\left[ f\right] \left( x\right)
=\int _{0}^{+\infty }W_{t}\left( x,y\right) f\left( y\right) dy,
\label{5.9}
\end{equation}
where
\begin{equation}
W_{t}\left( x,y\right) =\frac{2\sqrt{xy}e^{-t}}{(1-e^{-2t})}I_{\alpha
}\left( \frac{2xye^{-t}}{1-e^{-2t}}\right) \exp \left( -\frac{1}{2}\left(
x^{2}+y^{2}\right) \frac{1+e^{-2t}}{1-e^{-2t}}\right) .  \label{5.10}
\end{equation}
Here $I_{\alpha }\left( .\right) $ denotes the modified Bessel function of
the first kind and order $\alpha $ (\cite[p.66]{17}).

\section{Husimi's Q-function attached to $L_{\protect\alpha }$}

For $\sigma >1$\ and $m=0,1,...,\left[ \left( \sigma -1\right) /2\right] .$
A class of generalized negative binomial states (GNBS) attached to the
isotonic oscillator $L_{\alpha }$ can be defined by setting
\begin{equation}
\mid z\text{ },\sigma ,m,\alpha >=\left( \mathcal{N}_{\sigma ,m}\left(
z\right) \right) ^{-\frac{1}{2}}\sum\limits_{k=0}^{+\infty }\Phi
_{k}^{\sigma ,m}\left( z\right) \mid \psi _{k}^{\alpha }>  \label{6.1}
\end{equation}
where $\mathcal{N}_{\sigma ,m}\left( z\right) $\ is the factor in \eqref{4.10}, $\left\{ \Phi _{j}^{\sigma ,m}\left( z\right) \right\} $ are
defined in \eqref{3.8} and $\left\{ \mid \psi _{k}^{\alpha
}>\right\} $ are the Fock vectors given in \eqref{5.4}. The
diagonal representation of $e^{-t\mathbf{L}_{\alpha }}$ in the GNBSs
representation is now precised as follows.

 \begin{definition}\label{Def6.1} The Husimi's Q-function attached to the
operator $\mathbf{L}_{\alpha }$  is given through the mean value 
\begin{equation}
Q_{m}^{t}\left( \mathbf{L}_{\alpha }\right) \left( z\right) =\mathbb{E}%
_{\left\{ \mid z\text{ },\sigma ,m,\alpha >\right\} }\left( e^{-t\mathbf{L}%
_{\alpha }}\right) =<z ,\sigma ,m,\alpha \mid e^{-t\mathbf{L}_{\alpha
}}\mid z\text{ },\sigma ,m,\alpha >  \label{6.2}
\end{equation}
with respect to the set of GNBSs defined in \eqref{6.1}.
\end{definition}

 \begin{proposition}\label{Prop6.1} The mean value defined in \eqref{6.2}
 has the \ following expression
\begin{align}
\mathbb{E}_{\left\{ \mid z ,\sigma ,m,\alpha >\right\} } & \left( e^{-t%
\mathbf{L}_{\alpha }}\right) =\frac{\pi \left( 1-z\overline{z}\right)
^{\sigma }e^{-\left( 2\alpha +1\right) t}}{(\sigma -2m-1)}\left( \frac{%
\left( z\overline{z}-e^{-t}\right) \left( 1-z\overline{z}e^{-t}\right) }{%
\left( 1-z\overline{z}\right) ^{2}}\right) ^{m}
\nonumber \\
& \times \left( \frac{1-z\overline{z}}{1-z\overline{z}e^{-t}}\right) ^{\sigma
}P_{m}^{\left( \sigma -2m,0\right) }\left( 1+\frac{2e^{-t}\left( 1-z%
\overline{z}\right) ^{2}}{\left( z\overline{z}-e^{-t}\right) \left( 1-z%
\overline{z}e^{-t}\right) }\right) .  \label{6.3}
\end{align}
 \end{proposition}

 \noindent\textbf{Proof}. We start by inserting the expression \eqref{5.7}
of the operator $e^{-t\mathbf{L}_{\alpha }}$ into the equation \eqref{6.2} 
in which we also replace the GNBS by their definition in \eqref{6.1}. We obtain successively
\begin{align}
Q_{m}^{t}\left( \mathbf{L}_{\alpha }\right) \left( z\right)
&=\sum\limits_{k=0}^{+\infty }\exp \left( -t\lambda _{k}^{\alpha }\right) <z%
\text{ },\sigma ,m,\alpha \mid \psi _{k}^{\alpha }><\psi _{k}^{\alpha }\mid z%
\text{ },\sigma ,m,\alpha >  \label{6.4}
\\
&=\sum\limits_{k=0}^{+\infty }\exp \left( -t\lambda _{k}^{\alpha }\right)
\left| <z\text{ },\sigma ,m,\alpha \mid \psi _{k}^{\alpha }>\right| ^{2}
\label{6.5}
\\
&=\sum\limits_{k=0}^{+\infty }\exp \left( -t\lambda _{k}^{\alpha }\right)
\left| \left( \mathcal{N}_{\sigma ,m}\left( z\right) \right) ^{-\frac{1}{2}%
}\Phi _{k}^{\sigma ,m}\left( z\right) \right| ^{2}  \label{6.6}
\\
&=\left( \mathcal{N}_{\sigma ,m}\left( z\right) \right)
^{-1}\sum\limits_{k=0}^{+\infty }\exp \left( -t\left( 2\alpha +k+1\right)
\right) \left| \Phi _{k}^{\sigma ,m}\left( z\right) \right| ^{2}  \label{6.7}
\\
&=\left( \mathcal{N}_{\sigma ,m}\left( z\right) \right) ^{-1}e^{-\left(
2\alpha +1\right) t}\sum\limits_{k=0}^{+\infty }\left( e^{-t}\right)
^{k}\Phi _{k}^{\sigma ,m}\left( z\right) \overline{\Phi _{k}^{\sigma
,m}\left( z\right) }.  \label{6.8}
\end{align}
Now, to calculate the sum in \eqref{6.8} we make use of the
expression of the functions $\Phi _{k}^{\sigma ,m}\left( z\right) $ in \eqref{3.8} 
involving Jacobi polynomials and we apply the identity $%
\left( 4.8\right) $. This allows us to obtain the expression
\begin{align}
Q_{m}^{t}\left( \mathbf{L}_{\alpha }\right) \left( z\right)
&=\left( \mathcal{%
N}_{\sigma ,m}\left( z\right) \right) ^{-1}e^{-\left( 2\alpha +1\right)
t}\left( \frac{\left( z\overline{z}-e^{-t}\right) \left( 1-z\overline{z}%
e^{-t}\right) }{\left( 1-z\overline{z}\right) ^{2}}\right) ^{m}  \label{6.9}
\\
&\times \left( \frac{1-z\overline{z}}{1-z\overline{z}e^{-t}}\right) ^{\sigma
}P_{m}^{\left( \sigma -2m,0\right) }\left( 1+\frac{2e^{-t}\left( 1-z%
\overline{z}\right) ^{2}}{\left( z\overline{z}-e^{-t}\right) \left( 1-z%
\overline{z}e^{-t}\right) }\right) . \nonumber
\end{align}
Finally, we replace the factor $\mathcal{N}_{\sigma ,m}\left( z\right) $ by
its expression in \eqref{4.10}. \fin
\\

In the following we will use the $Q_{m}-$function presented above in order
to write an inequality involving  the thermodynamical potential associated
with the operator $\mathbf{L}_{\alpha }.$ This potential reads
\begin{equation}
\Omega _{\alpha }:=\frac{-1}{\beta }Tr\left( Log\left( 1+e^{-\beta \left(
\mathbf{L}_{\alpha }-\eta \right) }\right) \right)   \label{6.10}
\end{equation}
where $\eta $ is the chemical potential and $\beta =1/k_{B}T,$ $k_{B}$ is
the Boltzman constant and $T$ denotes the temperature. Let us put $\epsilon
=e^{\beta \eta }>0$ and state the following inequality.

 \begin{proposition}\label{Prop6.2}  Let $\sigma >1.$  Then, the
thermodynamical potential in \eqref{6.10} satisfy the
inequality
\begin{equation}
\max_{m\in \mathbb{Z}_{+}\cap \left[ 0,\left( \sigma -1\right) /2\right] }%
\left[ \frac{1}{\beta }\int _{\mathbb{D}}Log\left( \frac{1}{1+\epsilon
Q_{m}^{t}\left( \mathbf{L}_{\alpha }\right) \left( z\right) }\right) d\mu
_{\sigma ,m}\left( z\right) \right] \leq \Omega _{\alpha }  \label{6.11}
\end{equation}
for every $\beta >0.$
 \end{proposition}

 \noindent \textbf{Proof}. The form of \ the potential $\Omega _{\alpha }$ in \eqref{6.10} 
 suggests us to consider the function
\begin{equation}
\phi _{\epsilon }\left( u\right) =-Log\left( 1+\epsilon u\right) .
\label{6.12}
\end{equation}
So that we can rewrite \eqref{6.10} as
\begin{equation}
t\Omega _{\alpha }=Tr\left( \phi _{\epsilon }\left( e^{-t\mathbf{L}_{\alpha
}}\right) \right) ,  \label{6.13}
\end{equation}
where $\beta =t\in  \mathbb{R}_{+}$. We now apply the Berezin-Lieb
inequality \eqref{2.7} for the lower symbol $Q_{m}$ of the operator
$e^{-t\mathbf{L}_{\alpha }}$ in the GNBSs representation \eqref{6.2} to obtain the following inequality
\begin{equation}
\int _{\mathbb{D}}\left[ \phi _{\epsilon }\circ Q_{m}^{t}\left(
\mathbf{L}_{\alpha }\right) \right] \left( z\right) d\mu _{\sigma ,m}\left(
z\right) \leq Tr\left( \phi _{\epsilon }\left( e^{-t\mathbf{L}_{\alpha
}}\right) \right)  . \label{6.14}
\end{equation}
Making use of \eqref{6.12} and replacing the right hand side of 
\eqref{6.14} by $t\Omega _{\alpha }$ as in \eqref{6.13}, we
get an inequality that holds for every $m=0,1,...,\left[ \left( \sigma
-1\right) /2\right] .$ Therefore, we consider the maximum with respect to
the integer $m$ of the quantity in the left hand side of \eqref{6.14} in order to be close as possible to the value of $\Omega _{\alpha }$. \fin

\end{document}